\documentclass[letterpaper,english,english,prd,preprintnumbers,nofootinbib,showpacs,shortbibliography,notitlepage]{revtex4-1}
\usepackage[T1]{fontenc}
\usepackage[latin9]{inputenc}
\usepackage{amsmath}
\usepackage{amssymb}

\makeatletter

\usepackage{verbatim}
\usepackage{graphicx}
\usepackage{babel}
\makeatother

\usepackage{babel}
\begin{document}
\global\long\def\edge#1{\left.#1\right|}

\global\long\def\d{\mathrm{d}}
\global\long\def\yi{\varphi}

\global\long\def\bra#1{\left\langle #1\right|}
 \global\long\def\ket#1{\left| #1 \right\rangle }

\global\long\def\tr{\mbox{Tr}}

\global\long\def\order#1{\mathcal{O}\left(#1\right)}

\preprint{Alberta Thy 12-17}

\title{Top-quark loops and the muon anomalous magnetic moment}

\author{Andrzej Czarnecki }

\affiliation{Department of Physics, University of Alberta, Edmonton, Alberta,
Canada T6G 2E1}

\author{William J.~Marciano}
\email{marciano@bnl.gov}

\selectlanguage{english}

\affiliation{Department of Physics, Brookhaven National Laboratory, Upton, NY
11973, USA}
\begin{abstract}
The current status of electroweak radiative corrections to the muon
anomalous magnetic moment is discussed. Asymptotic expansions for
some important electroweak two loop top quark triangle diagrams are
illustrated and extended to higher order. Results are compared with
the more general integral representation solution for generic fermion
triangle loops coupled to pseudoscalar and scalar bosons of arbitrary
mass. Excellent agreement is found for a broader than expected range
of mass parameters. 
\end{abstract}
\maketitle

\section{Introduction}

The muon anomalous magnetic moment, $a_{\mu}=\left(g_{\mu}-2\right)/2$,
has been both precisely measured \cite{Bennett:2006fi} and very accurately
computed for the Standard Model (SM) \cite{Aoyama:2012wk,Jegerlehner:2009ry}.

Currently, there exists a provocative 3.5 sigma difference between
experiment and SM theory \cite{Davier:2017zfy}:
\begin{equation}
a_{\mu}^{\text{exp}}-a_{\mu}^{\text{SM }}=268\left(63\right){}_{\text{exp}}\left(43\right){}_{\text{SM }}\times10^{-11}=268\left(76\right)\times10^{-11}\label{eq1}
\end{equation}
which may indicate problems with the experiment and/or theory. A more
exciting possibility is that the discrepancy may be a harbinger of
\textquotedblleft New Physics\textquotedblright{} \cite{Czarnecki:2001pv},
beyond SM expectations. To clarify the situation, a more sensitive
experiment at Fermilab \cite{Grange:2015fou} is getting underway
with the goal of reducing the experimental uncertainty by a factor
of 4. Also, a distinctly low energy approach is being pursued at JPARC
\cite{Saito:2012zz}. Meanwhile, theoretical uncertainties, primarily
from hadronic loops, are expected to be further reduced (by perhaps
a factor of 2) from a combination of dispersion relations involving
$e^{+}e^{-}\to\text{hadrons}$ data and lattice gauge theory calculations
\cite{Lehner:2017kuc}. If \textquotedblleft New Physics\textquotedblright{}
is responsible for the current deviation, it should be fully exposed
with a solid $>5\sigma$ discovery during the next few years.

Given the importance of the theory calculations behind $a_{\mu}^{\text{SM }}=a_{\mu}^{\text{QED }}+a_{\mu}^{\text{hadronic }}+a_{\mu}^{\text{EW }}$,
it is important to scrutinize all of their underlying properties,
including the reliability of the computational methodology. Electroweak
(EW) Feynman loop diagrams contributing to $a_{\mu}^{\text{SM }}$
typically involve at least two mass scales: the muon mass and the
boson mass. At two loops, exact expressions for the resulting integrals
are complicated. In many cases they are not known. It is useful to
exploit the wide separation of these mass scales and expand the integrals
in their ratio. Such expansions have a long tradition in mathematical
physics \cite{bender1999methods}. In quantum field theory, they are
especially powerful when combined with dimensional regularization
that does not introduce additional scales (unlike for example Pauli-Villars
regularization). The crucial property is the vanishing of diagrams
that do not involve any mass scales, so called \emph{massless tadpoles},
analogous to the vanishing of scaleless integrals in the theory of
distributions \cite{GenFncGelfand}. 

In this paper, we provide a check on the asymptotic expansion method
used in the calculation \cite{CKM95,Czarnecki:1995sz} of the two-loop
electroweak (EW) contributions to $a_{\mu}^{\text{EW}}$. Those are
the corrections to the well known one loop contribution \cite{Jackiw:1972jz,Bars:1972pe,Fujikawa:1972fe,Bardeen72,Altarelli:1972nc}:
\begin{equation}
a_{\mu}^{\text{EW}}\left(\text{1 loop}\right)=\frac{5}{3}\frac{G_{\mu}m_{\mu}^{2}}{8\sqrt{2}\pi^{2}}\left[1+\frac{1}{5}\left(1-4\sin^{2}\theta_{W}\right)^{2}+\order{\frac{m_{\mu}^{2}}{M^{2}}}\right]=194.8\times10^{-11},\label{eq2}
\end{equation}
where $G_{\mu}=1.166\,378\,7\left(6\right)\times10^{-5}$ GeV$^{-2}$
is the Fermi coupling constant \cite{Marciano:1999ih,Mohr:2015ccw},
$m_{\mu}$ is the muon mass, $M$ represents the mass of electroweak
gauge or Higgs bosons, and $\theta_{W}$ is the weak mixing angle,
$\sin^{2}\theta_{W}=1-\frac{m_{W}^{2}}{m_{Z}^{2}}\simeq0.223$. Two
loop corrections are of the form:
\begin{equation}
a_{\mu}^{\text{EW}}\left(\text{2 loop}\right)=\frac{5}{3}\frac{G_{\mu}m_{\mu}^{2}}{8\sqrt{2}\pi^{2}}\sum_{i}C_{i}\frac{\alpha}{\pi}.\label{eq3}
\end{equation}
where in the 't Hooft-Feynman gauge the $C_{i}$ represent contributions
from about 240 Feynman diagrams as well as one loop induced counterterms.
Three loop leading log EW effects were shown to be negligible using
the renormalization group \cite{Degrassi:1998es, Czarnecki:2002nt}.
Collectively, for $m_{H}=125$ GeV, higher orders reduce $a_{\mu}^{\text{EW}}$
to \cite{Czarnecki:2002nt,Gnendiger:2013pva}
\begin{equation}
a_{\mu}^{\text{EW}}=a_{\mu}^{\text{EW}}\left(\text{1 loop}\right)+a_{\mu}^{\text{EW}}\left(\text{2 loop}\right)+a_{\mu}^{\text{EW}}\left(\text{3 loop leading logs}\right)=154\left(1\right)\times10^{-11}\label{eq4}
\end{equation}
where the uncertainty stems mainly from light quark two loop triangle
diagrams and non-leading-log three loop effects. The central value
in that result has been rounded off from $153.7\times10^{-11}$, after
individual contributions were computed up to $\order{0.1\times10^{-11}}$.

\begin{figure}[htb]

\begin{center}
\includegraphics[width=50mm]{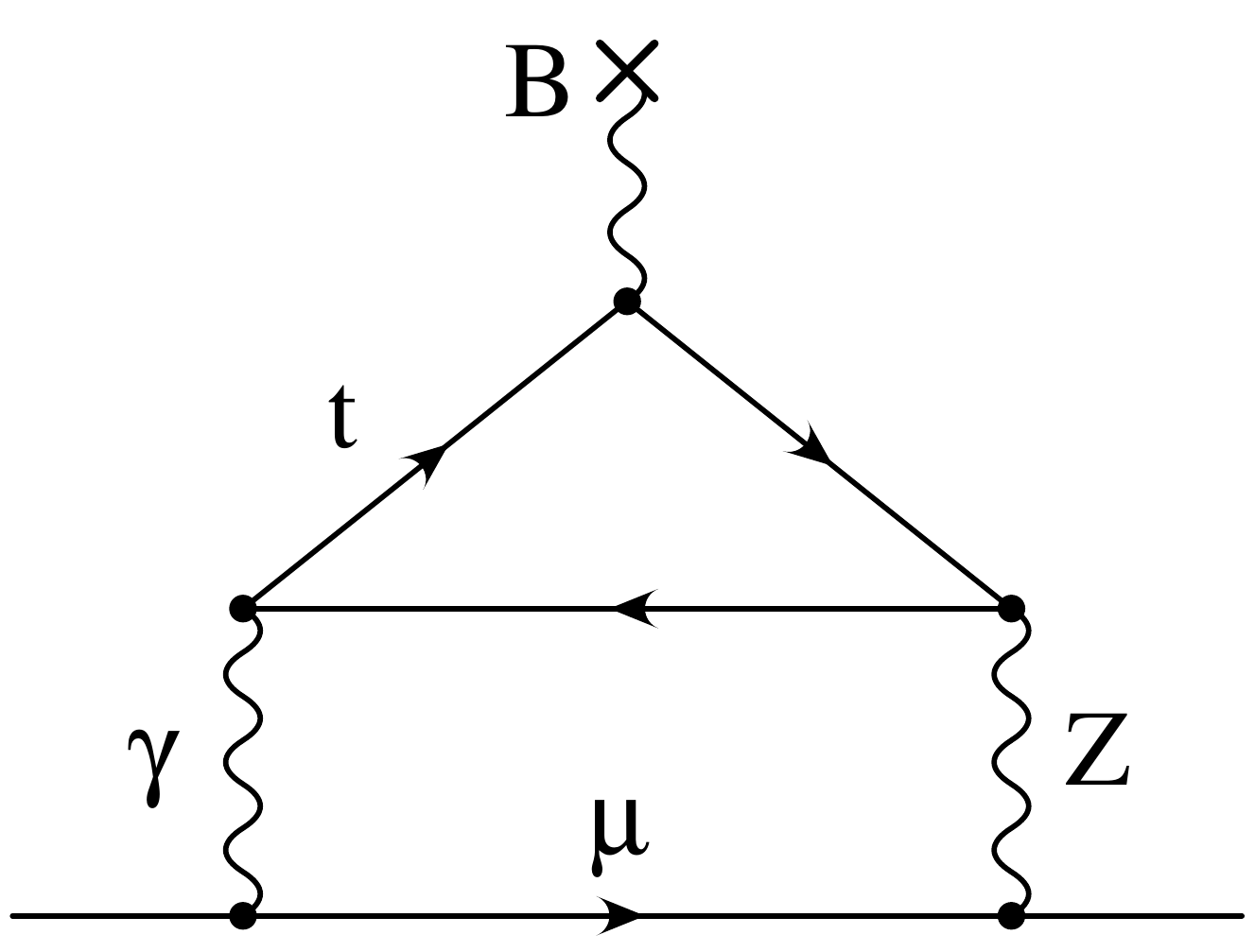}\hspace*{10mm}
\includegraphics[width=50mm]{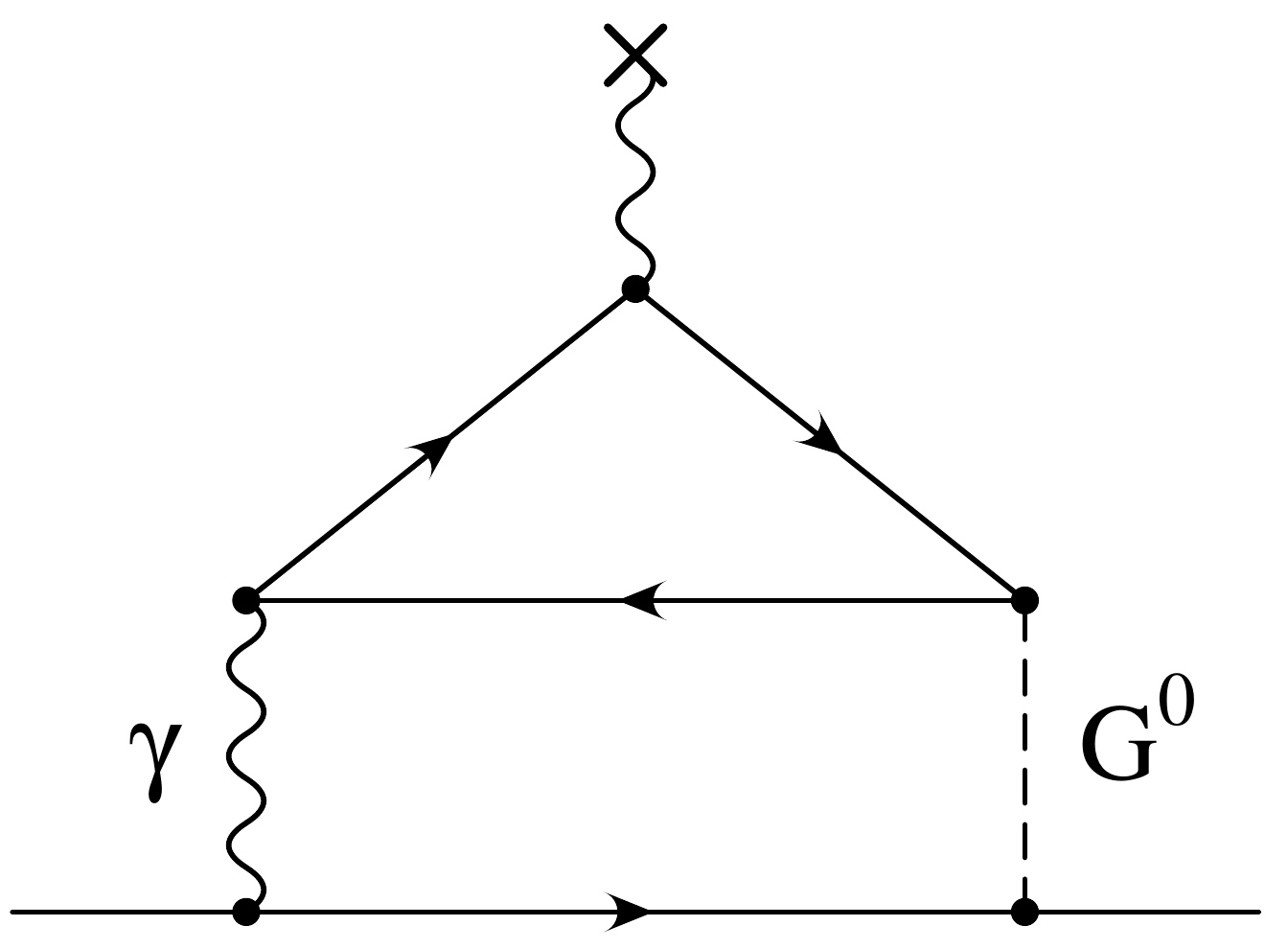}\hspace*{10mm}
\includegraphics[width=50mm]{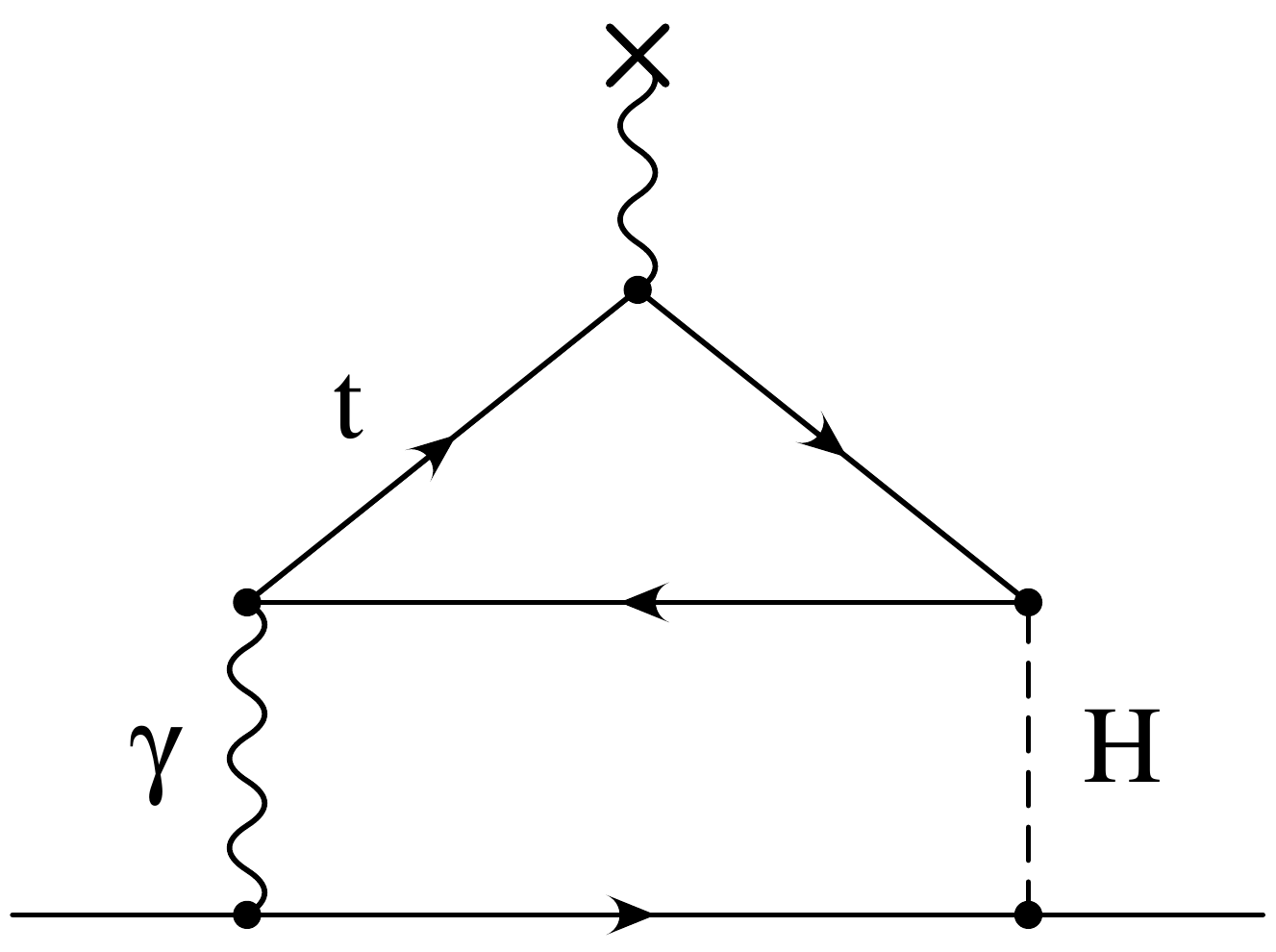}\\[3mm]
 (a) \hspace*{55mm} (b)\hspace*{55mm} (c)
\end{center}

\caption{Top-loop connected to the muon line by a photon and a $Z$ boson (a),
neutral Goldstone boson (b), Higgs boson (c). These diagrams have
additional versions: left-right reflected and with the external magnetic
field coupling to the other top line. \label{fig:Top-loop-connected}}

\end{figure}
To examine the convergence properties of the asymptotic expansions
used in eq.~\eqref{eq3}, we consider two particularly interesting
examples containing top quark triangle loops. The first case, illustrated
in Figs.~\ref{fig:Top-loop-connected}(a,b) represents a part of
the fermion anomaly diagrams. Together with contributions from the
$\tau$ lepton and $b$ quark they provide an anomaly-free
subset of two loop contributions. Light fermion effects are easily
calculated; but the top quark loop was originally computed as an asymptotic
expansion in $m_{Z}^{2}/m_{t}^{2}\simeq0.28$. We illustrate that
prescription below. Fig.~\ref{fig:Top-loop-connected}(c) represents
a Higgs scalar contribution for which the expansion parameter $m_{H}^{2}/m_{t}^{2}\simeq0.52$
is more of a concern, because of its relatively large value in comparison
with the underlying assumption $m_{H}^{2}/m_{t}^{2}\ll1$. Although,
as we will show, if several terms are included, the expansion remains
valid even for considerably larger values of $m_{H}^{2}/m_{t}^{2}$
than $0.52$. The diagrams in Figs.~\ref{fig:Top-loop-connected}(b)
and (c) are examples of what are called Barr-Zee diagrams in the literature
\cite{Barr:1990vd}. They often occur for heavy fermion triangle loops
coupled to heavy or light pseudoscalar or scalar particles. After
discussing top SM effects, we address the more general case of arbitrary
\textquotedblleft new physics\textquotedblright{} mass scales in similar
types of diagrams.

\section{Top loop diagrams}

As a concrete illustration of the asymptotic expansion method, we
begin by computing the diagrams shown in Fig.~\ref{fig:Top-loop-connected}(a,b)
using the 't Hooft-Feynman gauge. These diagrams involve three masses
whose squares we can consider widely-separated, 
\begin{equation}
m_{t}^{2}\gg m_{Z}^{2}\gg m_{\mu}^{2}.
\end{equation}
Indeed, with $m_{Z}=91.2$ GeV and $m_{t}=173$ GeV, the largest ratio
is $\frac{m_{Z}^{2}}{m_{t}^{2}}\simeq0.28$. Of course, the muon with
$m_{\mu}\simeq0.106$ GeV appears to be almost massless in comparison
with these heavy particles. However, its mass must be retained since
the electroweak correction $\Delta a_{\mu}$ vanishes in the massless
muon limit. Also, having three mass scales better illustrates the
power of the method. We thus have two small parameters, $\frac{m_{\mu}^{2}}{m_{Z}^{2}}\simeq10^{-6}$
and $\frac{m_{Z}^{2}}{m_{t}^{2}}$ in which the diagrams will be expanded.
We use the asymptotic expansion approach \cite{AE} to identify relevant
regions of loop momenta. The top-quark diagrams are an interesting
application of this approach with a relatively rich structure of the
hierarchy of momenta. 

\begin{figure}[h]
\begin{center}
\includegraphics[width=45mm]{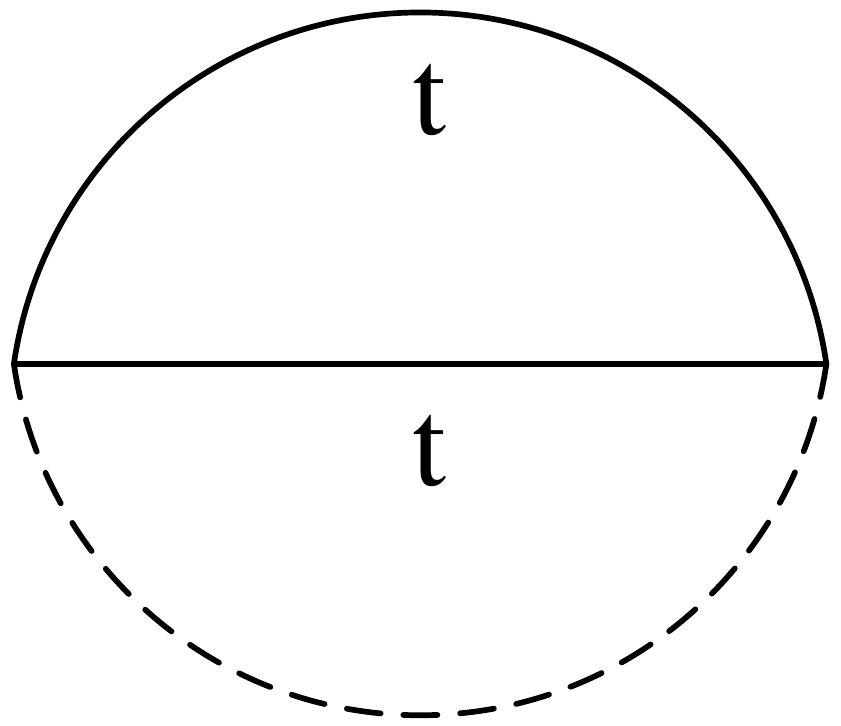}\hspace*{15mm}
\includegraphics[width=45mm]{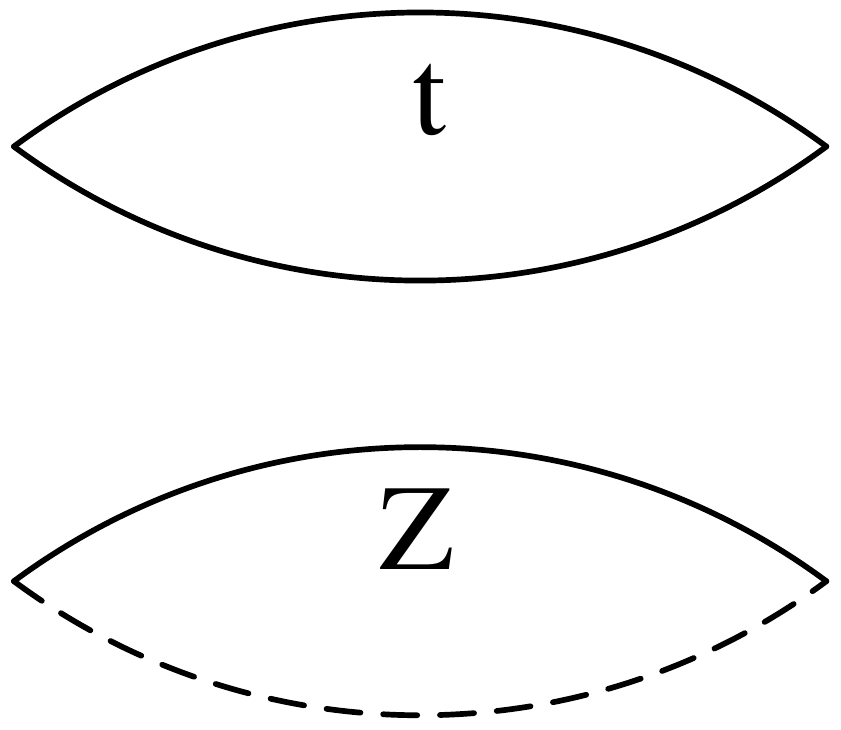}\hspace*{15mm}
\includegraphics[width=45mm]{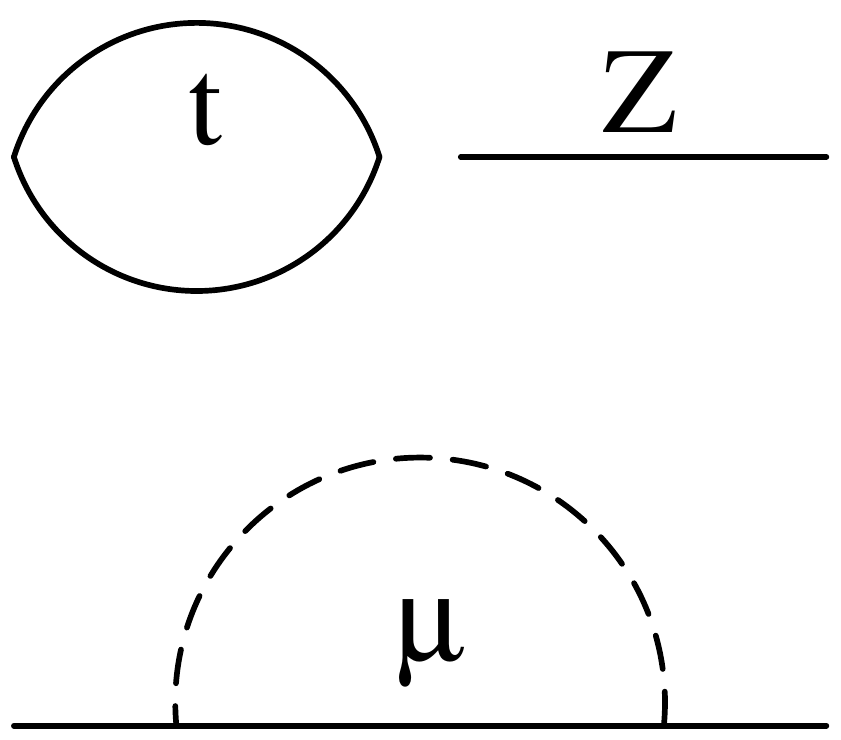}\\[3mm]
 (i) \hspace*{55mm} (ii)\hspace*{55mm} (iii)
\end{center}

\caption{Asymptotic expansion of the diagram in Fig. \ref{fig:Top-loop-connected}(a).
The momentum in the top quark loop is always on the order of $m_{t}$.
The other loop momentum is $\protect\order{m_{t}}$ in (i), $\protect\order{m_{Z}}$
in (ii), and $\protect\order{m_{\mu}}$ in (iii). Solid (dashed) lines
denote massive (massless) propagators.\label{fig:Asymptotic-expansion}}
\end{figure}

First, consider the case when both loop momenta are on the order of
the largest mass, $k_{1}\sim k_{2}\sim m_{t}$. Then the muon and
$Z$ propagators can be expanded around their massless limits and
the external muon momentum can also be treated as small. The resulting
integration is simplified because it depends only on the top-quark
mass, so the actual integrals that have to be computed are dimensionless
numbers. Their structure is shown in Fig.~\ref{fig:Asymptotic-expansion}(i):
the top-quark loop remains as in Fig.~\ref{fig:Top-loop-connected}(a),
but the $Z$ mass, the muon mass, and the external momentum are no
longer present (except as factors multiplying the integrand; they
can be taken out of the integration). Thus, the $Z$ propagator and
the muon propagator, together with the photon propagator, all form
some power of a massless propagator indicated in Fig.~\ref{fig:Asymptotic-expansion}(i)
by a dashed line. There are no external legs in that Figure because
external momenta are taken out of the integral. 

After the two-loop integration in this part, we find a divergent result.
Using dimensional regularization with $D=4-2\epsilon$,
\begin{equation}
\Delta C_\text{i}=\frac{m_{Z}^{2}}{m_{t}^{2}}\left[\frac{17}{15}-\frac{2}{5}\left(\frac{1}{\epsilon}-2\ln m_{t}^{2}\right)\right].\label{aDelta}
\end{equation}
Throughout this calculation we keep only results suppressed at most
by two powers of $\frac{1}{m_{t}}.$ In the 't Hooft-Feynman gauge,
the $Z$ diagrams are subleading, i.e.~suppressed by the inverse
top-quark mass. The leading contribution arises from the Goldstone
boson diagram depicted in Fig.~\ref{fig:Top-loop-connected}(b). 

The logarithm of the top mass in \eqref{aDelta} arises because $m_{t}$
is the only scale in this part of the integration. Since the integration
element is $\d^{D}k_{1}\d^{D}k_{2}$, the fractional power of mass
must be $m_{t}^{-4\epsilon}=1-2\epsilon\ln m_{t}^{2}+\order{\epsilon^{2}}$.
Multiplying the divergence, it gives $-\frac{2}{5\epsilon}\left(1-2\epsilon\ln m_{t}^{2}\right)=-\frac{2}{5\epsilon}+\frac{4}{5}\ln m_{t}^{2}$.
We anticipate that the divergences cancel in the sum of the three
regions shown in Fig.~\ref{fig:Asymptotic-expansion}. A top-quark
loop is present in all these regions, so a factor $m_{t}^{-2\epsilon}$
is present in all these three contributions. Thus we expect only half
of the $m_{t}$ logarithm in \eqref{aDelta} to survive in the final
result. Indeed, in eq.~(18) in \cite{CKM95} the logarithmic part
of the relevant diagram, denoted there by $\Delta C_{1\left(d\right)}^{\text{ferm}}\left(t\right)$,
is just $\frac{2}{5}\frac{m_{Z}^{2}}{m_{t}^{2}}\ln\frac{m_{t}^{2}}{m_{Z}^{2}}$.
We now proceed to show how the logarithm of $m_{Z}$ arises. 

To this end, consider the region where the integration momentum in
the lower loop in Fig.~\ref{fig:Top-loop-connected}(a) is $\order{m_{Z}}$.
Then we can still expand in the external muon momentum but must keep
the momentum dependence of the $Z$ propagator. On the other hand,
we can expand the top-quark propagators in the $Z$ momentum. The
result is a product of two one-loop integrals shown in Fig.~\ref{fig:Asymptotic-expansion}(ii):
for the top-loop integration, all external momenta including the other
loop momentum can be pulled out of the integral. The other loop depends
on the $Z$ boson mass as the only scale, so the muon propagator is
expanded in the external momentum and becomes again a part of the
massless line. We find that this region contributes
\begin{align}
\Delta C_\text{ii} & =\frac{m_{Z}^{2}}{m_{t}^{2}}\left[-\frac{7}{15}+\frac{2}{5}\left(\frac{1}{\epsilon}-\ln m_{Z}^{2}-\ln m_{t}^{2}\right)\right]+\frac{m_{\mu}^{2}}{m_{t}^{2}}\left[-\frac{34}{135}+\frac{4}{9}\left(\frac{1}{\epsilon}-\ln m_{Z}^{2}-\ln m_{t}^{2}\right)\right]\label{bDelta}
\end{align}
In this region, we observe a more complicated result, due to the dependence
on two mass scales. In the first term, the anticipated logarithm of
$m_{Z}$ appears, as well as the logarithm of the top mass with a
sign opposite than in\textbf{ \eqref{aDelta}}. In the sum of \eqref{aDelta}
and \eqref{bDelta}, the divergences multiplying $\frac{m_{Z}^{2}}{m_{t}^{2}}$
cancel and the logarithm $\frac{2}{5}\frac{m_{Z}^{2}}{m_{t}^{2}}\ln\frac{m_{t}^{2}}{m_{Z}^{2}}$
of eq.~(18) in \cite{CKM95} is reproduced. In addition, in \eqref{bDelta}
there is a term of order $\frac{m_{\mu}^{2}}{m_{t}^{2}}$. It is beyond
our intended accuracy but we keep it to further illustrate the generation
of logarithms of ratios of various scales. With this in mind, we proceed
to the remaining, third region (iii) in Fig.~\ref{fig:Asymptotic-expansion}.

In this region, the momentum in the lower loop in Fig.~\ref{fig:Top-loop-connected}(a)
is $\order{m_{\mu}}$. We can expand in it both the top-quark loop
and the $Z$ propagator, but we must retain the exact dependence of
the muon propagator on the external momentum. We find
\begin{equation}
\Delta C_\text{iii}=\frac{m_{\mu}^{2}}{m_{t}^{2}}\left[\frac{68}{135}-\frac{4}{9}\left(\frac{1}{\epsilon}-\ln m_{\mu}^{2}-\ln m_{t}^{2}\right)\right],\label{cDelta}
\end{equation}
a contribution that cancels the divergence in the second term in \eqref{bDelta}.
Summing $\Delta C_\text{i,ii,iii}$ we obtain
\begin{equation}
\Delta C_{Z}=\Delta C_\text{i}+\Delta C_\text{ii}+\Delta C_\text{iii}=\frac{m_{Z}^{2}}{m_{t}^{2}}\left(\frac{2}{3}+\frac{2}{5}\ln\frac{m_{t}^{2}}{m_{Z}^{2}}\right)+\frac{m_{\mu}^{2}}{m_{t}^{2}}\left(\frac{34}{135}-\frac{4}{9}\ln\frac{m_{Z}^{2}}{m_{\mu}^{2}}\right),\label{abcDeltaC}
\end{equation}
a finite result whose leading term reproduces eq.~(18) in \cite{CKM95}.
While the divergences canceled in the sum of the three regions, the
differences of logarithms containing various mass scales combined
to form logs of dimensionless ratios. 

The result in \eqref{abcDeltaC} must
be supplemented by the contribution of the neutral Goldstone boson,
Fig.~\ref{fig:Top-loop-connected}(b). The same three regions contribute
and we find, neglecting terms $\order{\frac{m_{\mu}^{2}}{m_{t}^{2}}}$,
the first two expansion terms in $m_{Z}^{2}/m_{t}^{2}$,
\begin{equation}
\Delta C_{G}=-\frac{16}{5}-\frac{8}{5}\ln\frac{m_{t}^{2}}{m_{Z}^{2}}+\frac{m_{Z}^{2}}{m_{t}^{2}}\left(-\frac{4}{9}-\frac{4}{15}\ln\frac{m_{t}^{2}}{m_{Z}^{2}}\right),\label{cGolds}
\end{equation}
where the leading terms reproduce eq.~(22) in \cite{CKM95},
while the subleading $\order{m_{Z}^{2}/m_{t}^{2}}$ terms are small
and were previously dropped. Here, we retain them and find, adding
eqs.~\eqref{abcDeltaC} and \eqref{cGolds}
\begin{equation}
\Delta C_{Z}+\Delta C_{G}=-\frac{16}{5}-\frac{8}{5}\ln\frac{m_{t}^{2}}{m_{Z}^{2}}+\frac{m_{Z}^{2}}{m_{t}^{2}}\left(\frac{2}{9}+\frac{2}{15}\ln\frac{m_{t}^{2}}{m_{Z}^{2}}\right)+\order{\frac{m_{Z}^{4}}{m_{t}^{4}},\frac{m_{\mu}^{2}}{m_{t}^{2}}}.
\end{equation}
Adding the $b$ and $\tau$ loops (taken from eq.~(17) in ref.~\cite{CKM95})
gives the contribution of the third generation \cite{Czarnecki:2002nt},
\begin{equation}
\Delta a_{\mu}^{\text{EW}}\left[\tau,b,t\right]=-\frac{\alpha}{\pi}\frac{G_{\mu}m_{\mu}^{2}}{8\pi^{2}\sqrt{2}}\left[\frac{8}{3}\ln\frac{m_{t}^{2}}{m_{Z}^{2}}-\frac{2}{9}\frac{m_{Z}^{2}}{m_{t}^{2}}\left(\ln\frac{m_{t}^{2}}{m_{Z}^{2}}+\frac{5}{3}\right)+4\ln\frac{m_{Z}^{2}}{m_{b}^{2}}+3\ln\frac{m_{b}^{2}}{m_{\tau}^{2}}-\frac{8}{3}\right]=-8.21\cdot10^{-11},\label{eq12}
\end{equation}
in agreement with the result first published in \cite{Czarnecki:2002nt}
and recently checked in an automated calculation \cite{Ishikawa:2017ouv}.
Note, although they play an insignificant role, contributing less
than $10^{-12}$ to eq.~\eqref{eq12}, we have retained terms of
order $m_{Z}^{2}/m_{t}^{2}$ in eq.~\eqref{bDelta} to illustrate
their effect on the asymptotic expansion.

\begin{figure}[htb]
\centering

\begin{tabular}{cc}
\includegraphics[width = 90mm]{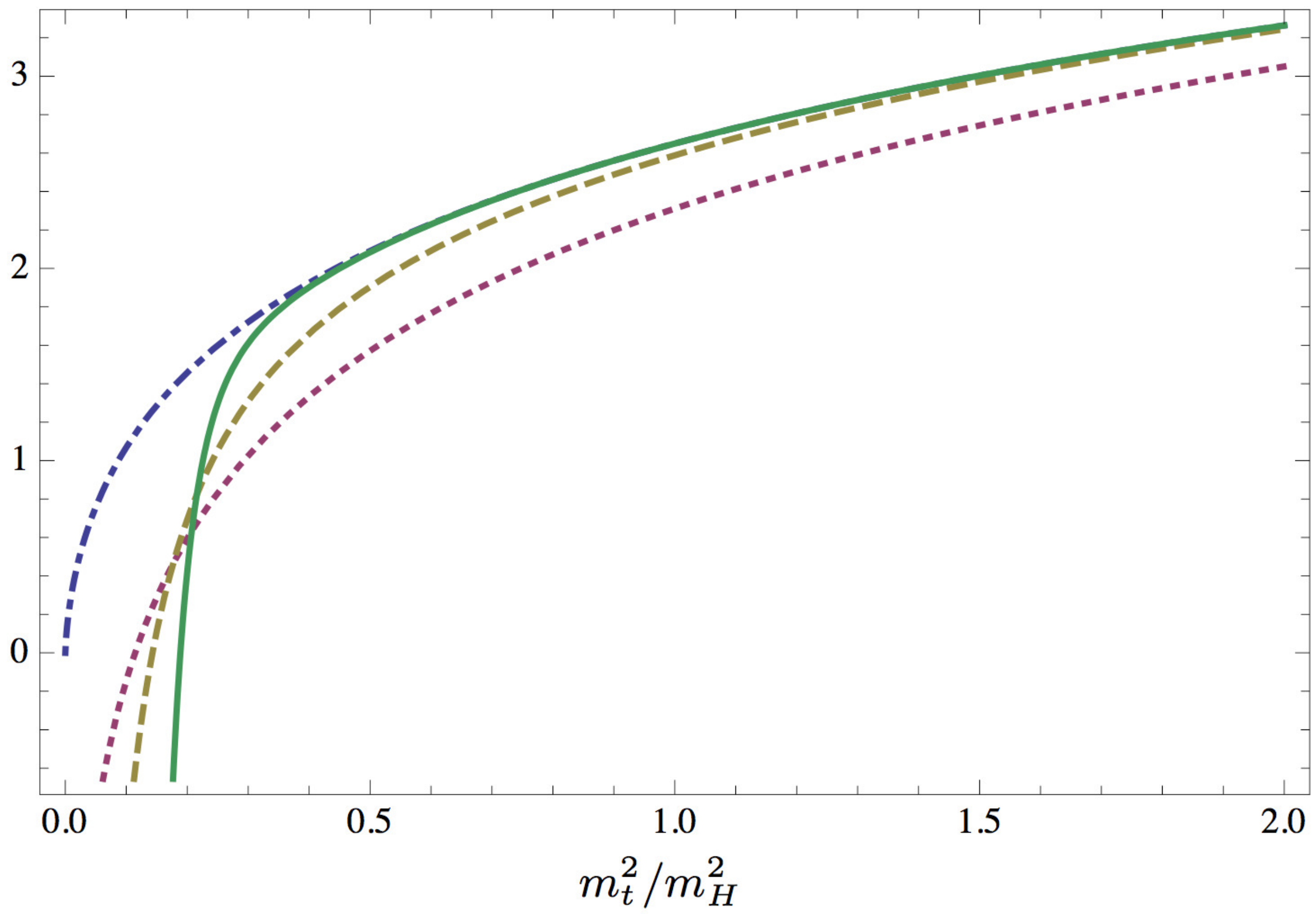}
\end{tabular}

\caption{Convergence of the asymptotic expansion of $-C_{H}$ on the basis
of eq.~\eqref{eq:Higgs}: solid line: all six powers of $m_{H}^{2}/m_{t}^{2}$;
dashed: just the first two powers; dotted: the first two terms only
(no terms $m_{H}^{2}/m_{t}^{2}$). The dot-dashed line shows the integral
representation in eq.~\eqref{intRepScalar}.\label{fig:Convergence}}

\end{figure}

For the analogous Higgs boson diagram, Fig.~\ref{fig:Top-loop-connected}(c),
we find after expanding to rather high order in $m_{H}^{2}/m_{t}^{2}$,
\begin{align}
\Delta C_{H} & =-\frac{104}{45}-\frac{16}{15}\ln\frac{m_{t}^{2}}{m_{H}^{2}}-\frac{m_{H}^{2}}{25m_{t}^{2}}\left(\frac{104}{15}+4\ln\frac{m_{t}^{2}}{m_{H}^{2}}\right)-\frac{m_{H}^{4}}{525m_{t}^{4}}\left(\frac{2692}{105}+16\ln\frac{m_{t}^{2}}{m_{H}^{2}}\right)\nonumber \\
 & -\frac{m_{H}^{6}}{315m_{t}^{6}}\left(\frac{971}{315}+2\ln\frac{m_{t}^{2}}{m_{H}^{2}}\right)-\frac{m_{H}^{8}}{5775m_{t}^{8}}\left(\frac{41758}{3465}+8\ln\frac{m_{t}^{2}}{m_{H}^{2}}\right)-\frac{m_{H}^{10}}{6435m_{t}^{10}}\left(\frac{267401}{90090}+2\ln\frac{m_{t}^{2}}{m_{H}^{2}}\right)-\dots\label{eq:Higgs}\\
 & =-3.2\text{ for }m_{H}^{2}/m_{t}^{2}=0.52.
\end{align}
Although $m_{H}^{2}/m_{t}^{2}=0.52$ is relatively large, higher order terms
in that expansion are suppressed by small coefficients. As a result,
the terms beyond the leading order contribute of order $10^{-12}$
to $a_{\mu}^{\text{EW}}$ which is covered by the uncertainty in eq.~\eqref{eq4}.
Expression \eqref{eq:Higgs} agrees with the integral representation
\cite{Barr:1990vd,Chang:2000ii,Cheung:2001hz},
\begin{equation}
\Delta C_{H}\left(z=\frac{m_{t}^{2}}{m_{H}^{2}}\right)=-\frac{8z}{5}\int_{0}^{1}\frac{1-2x\left(1-x\right)}{x\left(1-x\right)-z}\ln\frac{x\left(1-x\right)}{z}\d x.\label{intRepScalar}
\end{equation}
\begin{figure}[h]
\centering

\begin{tabular}{cc}
\includegraphics[width = 90mm]{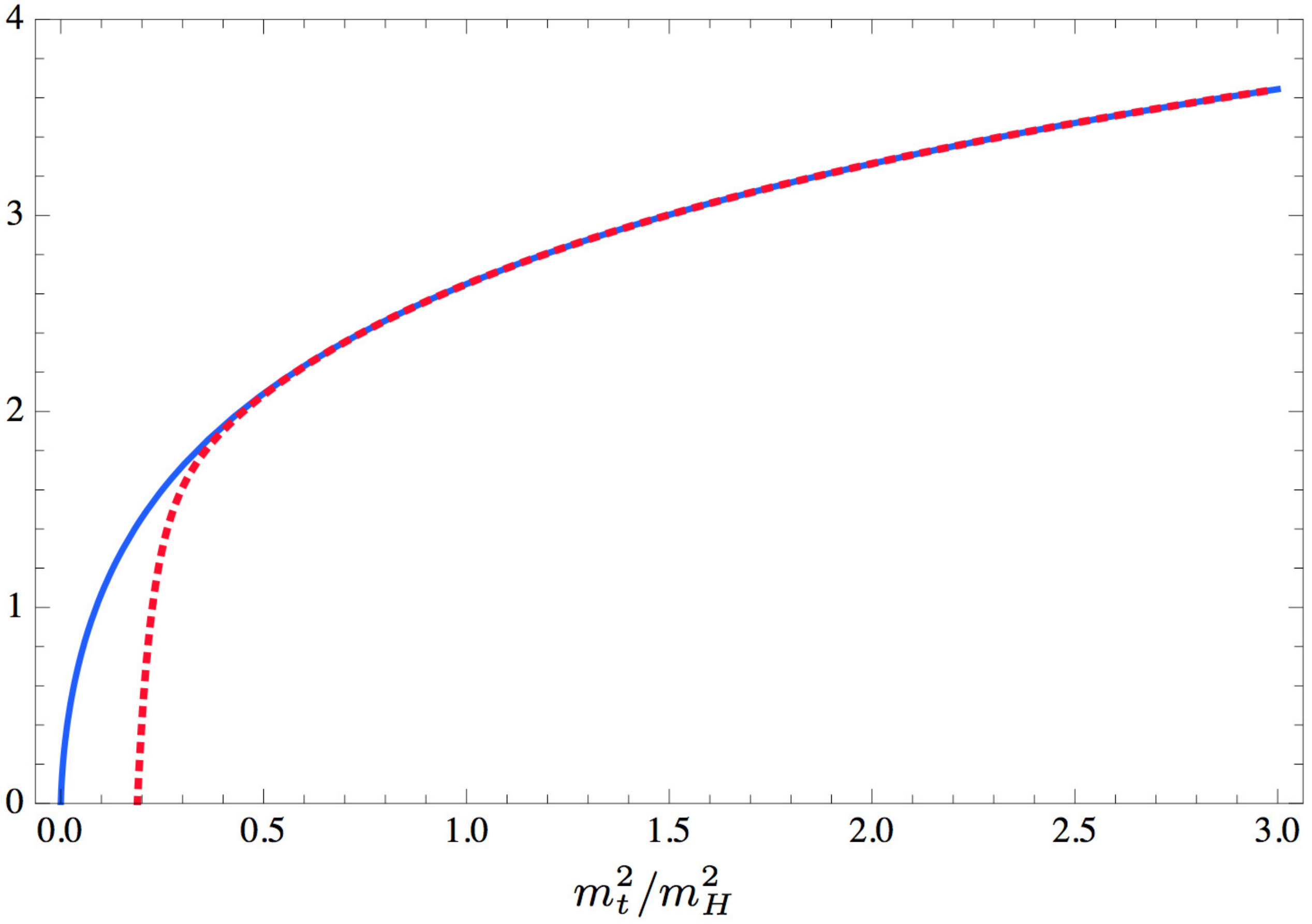} 
&
\includegraphics[width = 90mm]{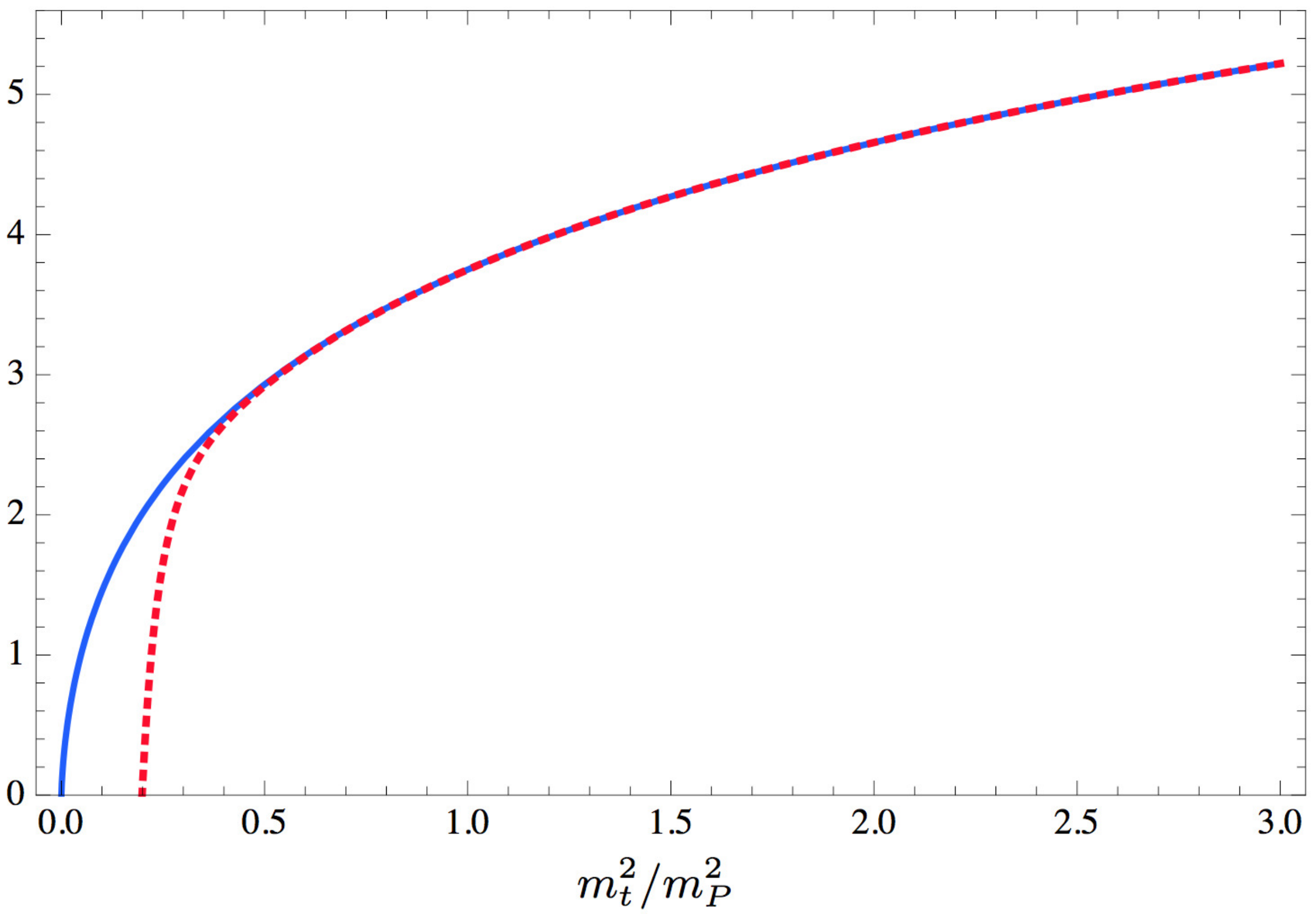} 
\\[2mm]
(a) & (b)
\end{tabular}

\caption{(a) Comparison of the asymptotic expansion in eq.~\eqref{eq:Higgs},
valid for $m_{t}>m_{H}$, (the dotted curve shows $-\Delta C_{H}$)
with the integral representation in eq.~\eqref{intRepScalar} (solid).
(b) Similar comparison for the Higgs boson replaced by a pseudoscalar
particle with mass $m_{P}$, the asymptotic expansion given by eq.~\eqref{asymPS}
and the integral representation by \eqref{intPS}.\label{fig:Comparison-with-int} }
\end{figure}
 Fig.~\ref{fig:Comparison-with-int}(a) shows that the asymptotic
expansion is valid in a remarkably broad range even for $m_{H}\simeq1.4m_{t}\simeq240$
GeV. Convergence of the expansion is illustrated with Fig.~\ref{fig:Convergence}.
For $m_{t}>m_{H}$, already the leading part of \eqref{eq:Higgs},
without $\frac{m_{H}^{2}}{m_{t}^{2}}$ corrections, differs from the
integral representation by only about 0.2. This corresponds to a contribution
to $\Delta a_{\mu}$ of about $10^{-12}$ which is in the noise but included in eq.~(\ref{eq4}) before roundoff.

If the scalar Higgs is replaced by a generic pseudo scalar with mass
$m_{P}$ and the same couplings as the $G^{0}$ in Fig.~\ref{fig:Top-loop-connected}(b),
the integral representation becomes \cite{Barr:1990vd,Chang:2000ii,Cheung:2001hz}
\begin{equation}
\Delta C_{P}\left(z=\frac{m_{t}^{2}}{m_{P}^{2}}\right)=-\frac{8z}{5}\int_{0}^{1}\frac{1}{x\left(1-x\right)-z}\ln\frac{x\left(1-x\right)}{z}\d x,\label{intPS}
\end{equation}
while our asymptotic expansion gives
\begin{equation}
\Delta C_{P}\left(z\right)=-\frac{16}{5}-\frac{8}{5}\ln z-\frac{\frac{20}{3}+4\ln z}{15z}-\frac{\frac{94}{15}+4\ln z}{75z^{2}}-\frac{\frac{319}{105}+2\ln z}{175z^{3}}-\frac{\frac{1879}{315}+4\ln z}{1575z^{4}}-\frac{\frac{20417}{6930}+2\ln z}{3465z^{5}}-\dots,\label{asymPS}
\end{equation}
The leading and next to leading terms agree with eq.~\eqref{cGolds}.
Again, as illustrated in Fig.~\ref{fig:Comparison-with-int}(b),
the asymptotic expansion agrees remarkably well with the integral
representation over a much broader range of parameters than one might
have expected. The quality of even the truncated asymptotic expansion
is nicely demonstrated in Fig.~\ref{fig:Convergence} where for $m_{t}^{2}/m_{H}^{2}>0.5$,
effects of higher order terms in $m_{H}^{2}/m_{t}^{2}$ are shown
to be relatively small.

The integral representations employed above have proved very useful
for confirming the validity of our asymptotic expansion for the muon
anomalous magnetic moment. In addition, they can be easily applied
to a broad range of single parameter ratios that may be required for
some \textquotedblleft New Physics\textquotedblright{} scenarios.
However, they are strictly valid in lowest order of the muon mass.
Asymptotic expansions can be extended to include powers and logarithms
of the muon mass, analogous to our example in eq.~\eqref{abcDeltaC},
where region (iii) was necessary to complete the determination of
the muon mass dependence. In addition, for multi loop diagrams involving
several mass ratios, the asymptotic expansion method is generally
applicable while relatively simple integral representations of the
type we used may be difficult to attain.

In summary, we have checked the numerical validity of the EW two loop
top quark triangle diagrams originally evaluated using asymptotic
expansions in $m_{Z}^{2}/m_{t}^{2}$ and $m_{H}^{2}/m_{t}^{2}$, by
comparing them with values obtained using the integral representation.
Agreement is excellent, even when retaining only one or two terms
in the expansion. Indeed, any truncation error is well below the uncertainty
budget of $\pm1\times10^{-11}$ assigned to $a_{\mu}^{\text{EW}}$
in eq.~\eqref{eq4}. Extending the expansion to six terms for a generic
scalar or pseudo scalar coupled to a heavy fermion of arbitrary mass
allowed us to compare the asymptotic expansion approach with the integral
representation for Barr-Zee diagrams that have been used for both
anomalous magnetic moment and electric dipole moment calculations.
We found asymptotic expansion agreement with the integral representation
for expansion parameters as large as 2 due to small coefficients in
the mass ratio squared expansion. That seems to indicate that the
real expansion parameter is actually much smaller than the mass ratio
squared and the asymptotic expansion method is very robust.

Acknowledgement: The work of A.~C.~was supported by Science and
Engineering Research Canada (NSERC). The work of W.~J.~M.~was supported
by the U.S. Department of Energy under grant DE-SC0012704.

%

\end{document}